\documentclass[10pt,aps,prd, floatfix,twocolumn,preprintnumbers,superscriptaddress,showpacs]{revtex4-1}
\usepackage{graphicx}
\usepackage{slashed}
\usepackage{amsmath,amsfonts,bm,bbold}
\usepackage{amssymb}
\begin{document}

\preprint{JLAB-THY-13-1757, July 24, 2013}

\author{A.~V.~Radyushkin}
\affiliation{Physics Department, Old Dominion University, Norfolk,
             VA 23529, USA}
\affiliation{Thomas Jefferson National Accelerator Facility,
              Newport News, VA 23606, USA
}
\affiliation{Bogoliubov Laboratory of Theoretical Physics, JINR, Dubna, Russian
             Federation}

\title{Sum Rules for Nucleon GPDs and Border Function Formulation}

\begin{abstract}

The  newly   developed  approach to model 
 nucleon generalized parton distributions (GPDs) $H$  and $E$   is 
based on two types of their    representation in terms of double distributions.
Within this approach, we 
  re-consider the derivation of GPD sum rules that 
 allow to use border functions
 $H(x,x)$ and $E(x,x)$
 instead of 
 full GPDs $H(x,\xi)$ and $E(x,\xi)$
  in the integrals producing Compton form factors
  of deeply virtual Compton scattering.   
  Using factorized DD Ansatz to model GPDs, we 
 discuss the relation between the border functions and 
 underlying 
 parton densities.  We find that a substantial contribution
 to $H(x,x)$ border function comes from   the extra term  required
 by new DD representations and related to  $E(x,\xi)$ GPD.

%\keywords{Generalized parton  distributions, Regge behavior, sum rules, analytic regularization }

\end{abstract}

\maketitle

\section{Introduction}

Building theoretical   models for   Generalized Parton Distributions
(GPDs)  \cite{Mueller:1998fv,Ji:1996ek,Radyushkin:1996nd,Collins:1996fb}
(for reviews see 
\cite{Ji:1998pc, Radyushkin:2000uy, Goeke:2001tz, Diehl:2003ny, Belitsky:2005qn,Boffi:2007yc,Radyushkin:2013nsa}) 
is a rather complicated  task, since they should satisfy several nontrivial requirements  
such as polynomiality \cite{Ji:1998pc},  
positivity \cite{Martin:1997wy,Pire:1998nw,Radyushkin:1998es},
  hermiticity  \cite{Mueller:1998fv}, time reversal invariance \cite{Ji:1998pc}, etc., that 
  follow from general principles  of  quantum  field  theory. 
In particular, an efficient way to impose the  polynomiality property
(which states that $x^n$ moment of a GPD $H(x,\xi;t)$  must be  a polynomial in $\xi$ of the
order not  higher than $n+1$)   
 is  to construct 
GPDs   from Double Distributions (DDs) $F(\beta, \alpha;t)$ 
 \cite{Mueller:1998fv,Radyushkin:1996nd,Radyushkin:1996ru,Radyushkin:1998es}. 
(Another way to satisfy  the 
 polynomiality   condition   is  to use  ``dual  parameterization''
 \cite{Polyakov:2002wz,Polyakov:2007rw,Polyakov:2007rv,SemenovTianShansky:2008mp,Polyakov:2008aa}).  
 
 In the course of  development of GPD theory, it was realized that
 the simplest $F(\beta, \alpha;t) \to H(x,\xi;t)$
  reconstruction method \cite{Radyushkin:1997ki}, derived from 
 the analysis 
 of scalar composite operators in scalar field theories, 
 does not produce the highest, $(n+1)^{\rm st}$ 
 power of the skewness parameter $\xi$ that is required 
 for vector operators.  For pion GPDs, to handle this problem, 
 a   parametrization  involving two DDs was formulated
\cite{Polyakov:1999gs}, with the second DD $G(\beta, \alpha)$
capable of generating the required $\xi^{n+1}$ power. 
It was also proposed \cite{Polyakov:1999gs} to use  a  ``DD plus D''
 decomposition, 
in which the second DD $G(\beta, \alpha)$  is reduced to a function $D(\alpha)$
of one variable,  the $D$-term, that is   solely  responsible for the $\xi^{n+1}$ contribution.
Later, it was emphasized \cite{Teryaev:2001qm} that the 
two-DD description is redundant
(which is natural since the pion is described by just one GPD
$H(x,\xi;t)$): the total contribution is determined 
by $\partial F(\beta,\alpha)/\partial \beta +\partial G(\beta,\alpha)/\partial \alpha$
combination,  and 
one can reshuffle terms between
$F$ and $G$ provided that this sum is unchanged 
(this is analogous to performing a  ``gauge transformation'' \cite{Teryaev:2001qm}).
The choice $G (\beta,\alpha) = \delta (\beta) D(\alpha)$ 
may be called  ``Polyakov-Weiss'' gauge.
  In this gauge, the 
non-D-term part is given by just one function
$F_D (\beta,\alpha)$. Another choice, ``one-DD'' gauge 
corresponds to $F_O(\beta,\alpha) = \beta {\cal F} (\beta,\alpha)$,
$G_O(\beta,\alpha)= \alpha  {\cal F} (\beta,\alpha)$.
The combination of $F$ and $G$ corresponding 
to the one-DD gauge naturally appears for matrix elements 
of twist-2 vector operators \cite{Belitsky:2000vk}. 

For  nucleon, a straightforward parametrization
involves 3 double distributions  \cite{Polyakov:1999gs,Tiburzi:2004mh,Mezrag:2013mya}, one of which is redundant (there are only two GPDs, $H(x,\xi;t)$
and $E(x,\xi;t)$),
i.e., as in pion case, one can perform gauge transformations that
do not change the total sum. Again, 
imposing symmetrization of indices involved in the definition
of local twist-2 operators, one obtains 
a natural parametrization
in terms of just two DDs ${\cal A}$ and ${\cal B}$ \cite{Belitsky:2005qn}.

Within the DD approach,  the problem of  constructing a  model  for a  GPD 
 converts into a  problem of  building a  model
 for the relevant DD. The advantage of using DDs as a starting point 
 (apart from satisfying the polynomiality condition) is 
a  simple physical picture they imply: 
 $F(\beta, \alpha;t)$   behave  like 
 usual parton distribution functions (PDFs)  with respect to 
 its variable $\beta$ and   as a meson 
 distribution amplitude  (DA) with respect to $\alpha$
 (and also  as  a form  factor 
 with respect to the invariant  momentum transfer $t$).

 The {\it factorized DD ansatz} (FDDA) 
\cite{Radyushkin:1998es,Radyushkin:1998bz} proposes 
to  build a  model DD $F(\beta, \alpha)$  
(in the simplified formal $t=0$  limit) as  a product of the usual 
parton density  $f(\beta)$
and a  profile function $h(\beta, \alpha)$  that  has an $\alpha$-shape of a 
meson DA.  Given the ambiguity of DDs involved,
one should decide, in which gauge the FDDA is applied.
Originally, FDDA in case of pion  was used 
for the ``DD+D'' decomposition, or in Polyakov-Weiss gauge.  
A  pion GPD model  based on one-DD gauge 
was built in our paper  \cite{Radyushkin:2011dh}. 
The complication is that applying FDDA to a one-DD representation
one should reconstruct GPD from $f(\beta)/\beta$.
The extra $1/\beta$ factor combined with the Regge $\beta^{-a}$
singularity of the parton density  $f(\beta)$ results 
in a non-integrable singularity for $\beta=0$.
In  our  construction  \cite{Radyushkin:2011dh}
(see also \cite{Radyushkin:2012gba}), we 
 separated   DD ${\cal F} (\beta,\alpha)$ in the 
``plus'' part $[f(\beta,\alpha)]_+$
that gives zero after integration over $\beta$,
and the $D$-term part $\delta (\beta) D(\alpha)/\alpha$. 
For DDs  singular in small-$\beta$ region, such a separation
serves also as a renormalization prescription
substituting a formally divergent integral over $\beta$
by ``observable'' $D$-term.

For the nucleon, the analog of ``DD+D'' construction has the structure
$H_{DD} +D$ for  GPD $H$ and $E_{DD} -D$ for GPD  $E$.
In fact, the combination that has the simplest  DD representation
is  $A=H+E$,  whereas  $B=-E$ has a DD representation identical in structure
to one-DD representation of the pion case \cite{Belitsky:2005qn,Radyushkin:2013hca}.
Using FDDA, one would reconstruct $A$ from the forward function
$f(x)+e(x)$ (which, roughly speaking,  is equivalent to 
taking $H+E=H_{DD}+E_{DD}$),  while $B$ is reconstructed from $-e(x)/x$,
which requires special treatment of the $x=0$ singularity. 
The result for $E$ may be written as \mbox{$E= E_{DD}+\xi E_+^1-D$} 
\cite{Radyushkin:2013hca},
where $E_+^1(x,\xi)$ is an extra  term specific to the one-DD parametrization.
Unlike the \mbox{$D$-term,}  it does not vanish at the border point $x=\xi$. 
As a result,  $H= H_{DD} - \xi E_+^1 +D$, i.e. 
GPD $H$ also acquires an extra term affecting its value at border 
point $x= \xi$.

The importance of knowing GPDs at the border point 
was realized from the earliest papers on 
deeply virtual Compton scattering (DVCS) \cite{Ji:1996ek,Ji:1996nm}.
In particular,  
 the imaginary part of the Compton amplitude ${\cal C} (\xi)$,  at  the leading order 
given by $H(\xi,\xi)$, determines the 
magnitude of the single-spin asymmetry \cite{Ji:1996ek,Ji:1996nm} and, hence, 
 is directly measurable in DVCS experiments.
However, the real part of  the leading order  ${\cal C} (\xi)$ 
(measurable through DVCS cross section)  is given by the 
principle value (PV)  integral of $H(x,\xi)/(x- \xi)$ and, thus,
apparently requires to know $H(x,\xi)$ for all range of $x$ 
and $\xi$ values. But, from the dispersion relation considerations \cite{Teryaev:2005uj,Anikin:2007yh},
 it was argued  that one can substitute $H(x,\xi)$ by  the border function 
 $H(x,x)$ in  the PV integral (one should  also add  a subtraction constant 
 $\Delta$ determined by the D term). 
 Thus, to calculate the whole leading-order Compton amplitude ${\cal C} (\xi)$
 (or form factor ${\cal C} (\xi;t)$, if the $t$-dependence is taken care of)
it is sufficient to know  the border functions, i.e., GPDs for $x=\xi$. 
 In pion case, the relation that allows to substitute $H(x,\xi)$ by $H(x,x)$
 in the integral for Compton amplitude (``GPD sum rule'') 
 was established \cite{Teryaev:2005uj,Anikin:2007yh}  using 
 double distribution representation for GPDs.
 For the nucleon case, the derivation was done \cite{Diehl:2007jb} 
 assuming ``$H_{DD} +D$, $E_{DD} -D$'' decomposition.
 As argued in our paper \cite{Radyushkin:2013hca}  and outlined above, the actual 
 decomposition in the nucleon case is somewhat more involved.  
 In the present paper, we  apply the technique of Ref.
\cite{Radyushkin:2013hca}  to rederive GPD sum rules
for   the nucleon case.   Another goal is to build 
simple (but  realistic) models for border functions 
of nucleon GPDs and analyze their properties.

The paper is organized as follows.
In Sect. II,  we give a brief  
 review of  GPD  sum rules that relate the difference 
 between full GPD $H(x,\xi)$ and border function $H(x,x)$ 
 with $D$ term. In particular, we present  the derivation of
 the GPD sum rule for  the simplest case that
 is similar to that originally given  in Refs. 
 \cite{Teryaev:2005uj,Anikin:2007yh,Diehl:2007jb}.
 In Sect. III, we discuss nucleon GPDs and DDs.
 Sum rules for nucleon GPDs are considered in Sect. IV.
 In Sect. V, we consider  ``secondary'' GPD sum rules for the nucleon,
 that correspond to formally taking $\xi=0$  limit 
 of original GPD sum rules.
 Models for the border functions of nucleon GPDs are
 considered in Sect. VI.
 In final  section, we summarize the results of the paper.

\section{Preliminaries: GPD sum rules}

\subsection{Formulation}

 The Compton amplitude describing DVCS is given by Compton form factors 
 which have the generic structure 
 \begin{align}
{\cal C} _H(\xi) = &   \int_{-1}^1  \frac{ H(x,\xi)}{x-\xi + i \epsilon }\,  dx  
\nonumber \\ &= 
-i \pi H(\xi,\xi) + 
 P\int_{-1}^1  \frac{H(x,\xi)}{x-\xi}\,  dx \ .
  \label{eq_SRG} 
 \end{align}
 It is well-known that the imaginary part   
 ${\rm  Im} \, {\cal C} (\xi)$ is given by the GPD  $H(x , \xi )$ on the diagonal 
$x=\xi$.  
From  dispersion relation considerations, the real part  ${\rm  Re} \, {\cal C} (\xi)$
for $\xi$ outside the $(-1,1)$ interval 
should also be  expressed, up to a subtraction constant, 
through an integral over the  imaginary part
 \begin{align}
{\rm  Re} \,  {\cal C}_H (\xi) = 
\int_{-1}^1  \frac{H(x,x)}{x-\xi}  \,  dx + {\rm const} \ .
  \label{eq:SRGRe} 
 \end{align}
 This implies that, for physical $\xi$,  the principal value integrals 
 of $H(x,\xi)$ and $H(x,x)$ with $1/(x-\xi)$ differ by a constant:
 \begin{align}
  P\int_{-1}^1  \frac{H (x, x) - H(x,\xi)}{x-\xi}\,  dx  =  \Delta  \ .
  \label{eq_SRGdel} 
 \end{align}
 This  expectation was confirmed  \cite{Teryaev:2005uj} using the formalism
 of double distributions. 
 
Namely, assume that   $H(x,\xi)$  is a sum of the ``DD'' part $H_{DD} (x, \xi)$ given by
 the double distribution (DD)  representation 
 \begin{align}
  {H_{DD} (x,\xi) }  =
  \int_{\Omega}    {\cal F}  (\beta,\alpha) \,  \delta (x - \beta -\xi \alpha)  \, d\beta \, d\alpha 
 \label{GPDF}
 \end{align}
 and the D term part ${\rm sgn} (\xi) D(x/\xi)$, with $D(x/\xi)$ vanishing
 for \mbox{$x= \pm \xi$} and outside the $|x|>|\xi|$ region. For   the DD part,  
 using Eq.~(\ref{GPDF})  one  has
\begin{align}  
P&  \int_{-1}^1  \frac{H_{DD} (x, x) -H_{DD}  (x,\xi)  }{x-\xi}\,  dx    \nonumber
 \\ &
 = 
 P \int_{-1}^1  \frac{dx}{x-\xi}\,    
 \int_{\Omega} {\cal F}  (\beta,\alpha)   \,
  \, d\beta \, d\alpha   
  \nonumber \\ & \times
  \biggl [ \delta (x(1-\alpha) - \beta ) -   \delta (x - \beta -\xi \alpha)  
     \biggr ] 
 \nonumber     \\ & 
     =   
P  \int_{\Omega} {\cal F}  (\beta,\alpha)   \,
  \, d\beta \, d\alpha   
    \nonumber \\ & \times
  \biggl [  \frac{1/(1-\alpha)}{\beta/(1-\alpha)  - \xi } - \frac {1 }{\beta- \xi (1-\alpha) }
     \biggr ] =0  \  . 
 \label{eq_SRG2xx0} 
\end{align}

Thus, as noticed in Refs.~\cite{Teryaev:2005uj,Diehl:2007jb}, seemingly different delta-functions have converted
$1/(x-\xi)$  into identical  expressions.    
As a result,  
\begin{align}
 \label{HDDSR}
 P\int_{-1}^1  \frac{H_{DD}  (x, x) }{x-\xi}\,  dx - P\int_{-1}^1  \frac{ H_{DD} (x,\xi)}{x-\xi}\,  dx  =  0  \  .
\end{align}

The  D term  gives   zero contribution into $H(x,x)$, while its contribution
into $H(x,\xi)$ produces 
 the claimed  constant 
  \begin{align}  
   {\rm sgn}(\xi)\, P \int_{-1}^1  \frac{  \theta (|x|<|\xi|)  \,  D(x/\xi) }{x-\xi}\,  dx 
 = -  \int_{-1}^1  \frac{D(\alpha)}{1-\alpha} \, d \alpha  \ .
  \label{SRD2nd}
    \end{align}
    As a result, 
 the constant  $\Delta$ is related to D  term by 
  \begin{align}
 \Delta  \equiv 
    \int_{-1}^1  \frac{D(\alpha)}{1-\alpha} \, d \alpha \ .
   \label{eq_SRD} 
  \end{align}
    
    Thus, the Compton form factor  ${\cal C}_H (\xi)$  is expressed 
    through a one-dimensional {\it border function} $H(\xi, \xi) \equiv {\mathfrak h} (\xi)$
     \begin{align}
{\cal C} _H (\xi) = 
-i \pi {\mathfrak h} (\xi) + 
 P\int_{-1}^1  \frac{{\mathfrak h} (x)}{x-\xi}\,  dx + \Delta \ .
  \label{eq_Cxi} 
 \end{align}

As discussed in  Ref.  \cite{Radyushkin:2013hca},  in the nucleon 
 case GPD $H$ is given by 
 a sum of two  terms that have  different-type  DD representations.
The first  of them  is  the same as in  Eq.(\ref{GPDF}), while  the other  has a  
  more singular  structure  similar to the 
 pion ``one-DD'' representation \cite{Radyushkin:2011dh}. 
 As a result, the structure of $H(x,\xi)$ GPD is more complicated 
 than it was assumed in the ``DD+D'' Ansatz. 
 One of the goals of  the present paper is to 
 reanalyze the sum rule within the framework  of 
 Ref.  \cite{Radyushkin:2013hca},  and 
 study to which extent  the original proofs still work.

\section{Nucleon GPDs}

\subsection{Definitions of DDs and GPDs}

In the  nucleon case, for unpolarized target,   one can parametrize  
 \begin{align}
&  \langle p' |   \bar \psi(-z/2) \slashed z   \, \psi (z/2)|p \rangle  |_{\rm twist-2} 
\label{pmurmu2}
 \\ & = 
\int_{\Omega}   e^{-i \beta (Pz) -i\alpha (rz)/2} \,  \Big \{
\bar u ( p')  \slashed z   \,  u ( p)\, {\cal A}  (\beta, \alpha) \nonumber
 \\  &
+ \frac{\bar u ( p')     u ( p) }{2 M_N} \, \big [ 2\beta (Pz) +\alpha (rz) \big ]   {\cal B}  (\beta, \alpha)   
\Big \}  \, d\beta \, d\alpha \ 
+{\cal O} (z^2)  
 \  .  \nonumber
\end{align}
Here, the functions ${\cal A,B}$ are  DDs corresponding to the 
combinations $A=H+E$ and $B= -E$
of usual  GPDs $H$ and $E$ (see Ref. \cite{Belitsky:2005qn}).
These GPDs  may be expressed in terms of  relevant DDs as
\begin{align}
 {A(x,\xi) }  =
 \int_{\Omega}    {\cal A}  (\beta,\alpha) \,  \delta (x - \beta -\xi \alpha)  \, d\beta \, d\alpha 
\label{GPDA}
\end{align}
and
\begin{align}
 {B(x,\xi) }  ={x}
 \int_{\Omega}    {\cal B}  (\beta,\alpha) \,  \delta (x - \beta -\xi \alpha)  \, d\beta \, d\alpha  \  .
\label{GPDB}
\end{align}

Notice that we have two different types of relations between
GPDs and DDs: $A(x, \xi)$ is obtained from
its DD ${\cal A} (\beta,\alpha)$  in a straightforward
``classic'' way of Eq.(\ref{GPDF}), while 
$B(x, \xi)$ is calculated from
 ${\cal B}(\beta,\alpha)$  using the formula
 with extra factor of $x$ involved 
(like in the ``one-DD''  representation  for the pion GPD,
see Refs. \cite{Radyushkin:2011dh,Radyushkin:2013hca}).
 The difference is due to
 the factor $[ 2\beta (Pz) +\alpha (rz)  ] $
 in the ${\cal B}$ part.

\subsection{Structure of DD representation} 

In the forward limit, we have
\begin{align}
A(x,0) =H(x,0)+E(x,0) = f(x) +e(x)
\end{align}
and 
\begin{align}
B(x,0) =-E(x,0) = -e(x)  \ .
\end{align}
The  first formula suggests the splitting 
\begin{align}
{\cal A} (\beta,\alpha) = 
{\cal F}  (\beta,\alpha) + {\cal E}_A  (\beta,\alpha)  \   ,
\end{align}
with ${\cal F}  (\beta,\alpha)$  and $ {\cal E} _A (\beta,\alpha)$
producing ``DD parts'' 
  \begin{align}
    F_{DD} (x,\xi) = \int_{\Omega}  {\cal F}  (\beta,\alpha)
    \,  \delta (x - \beta -\xi \alpha)    \, d\beta \, d\alpha
\label{GPDHf}
\end{align}
and 
\begin{align}
    E_{DD}  (x,\xi) = \int_{\Omega}  {\cal E}_A  (\beta,\alpha)
    \,  \delta (x - \beta -\xi \alpha)    \, d\beta \, d\alpha
\label{GPDE}
\end{align}
of GPDs  $H(x,\xi)$ and $E(x,\xi)$. 
The forward limit of    $B(x,\xi)/x$, i.e., $-E(x,0)/x$  is  obtained by integrating
 ${\cal B} (\beta,\alpha)$ over $\alpha$.
 This observation suggests the 
  representation
\begin{align}
{\cal B}  (\beta,\alpha) = 
- \frac{{\cal E}_B (\beta,\alpha)}{\beta} \  .
\end{align}
Since both ${\cal E}_A (\beta,\alpha)$ and ${\cal E}_B (\beta,\alpha)$
produce $e(x)$ in the forward limit, we have
\begin{align}
 \int_{-1}^1  [ {\cal E}_A  (\beta,\alpha)-{\cal E}_B (\beta,\alpha)]
\, d\alpha =0   \  .
\label{EaEb}
\end{align}
In our paper \cite{Radyushkin:2013hca}, the model 
${\cal E}_A (\beta,\alpha)={\cal E}_B (\beta,\alpha)$ was used. 
It should be emphasized, however,  that this is just a model assumption.
In particular, the same parton distribution in the  forward limit 
may be produced  by  DDs with different $\alpha$ profiles. 

The fact that, after integration over $\alpha$,
the DD   ${\cal B} (\beta,\alpha)$ gives
$-e(x)/x$ while ${\cal A} (\beta,\alpha)$  produces the combination 
$ f(x)+e(x)$ that  does not involve $1/x$ factor,
is an evidence that ${\cal B}  (\beta,\alpha)$ is more singular 
for small $\beta$ than ${\cal A}  (\beta,\alpha)$.
Because of possible singularity of  ${\cal B} (\beta,\alpha)$  for  $\beta=0$,
we write it in the ``DD$_++ D$''  representation:
\begin{align}
\label{BplusD}
{\cal B} (\beta,\alpha) = {{\cal B} _+(\beta,\alpha)}   
+
\delta (\beta) \frac{D(\alpha) }{\alpha}  \  ,
\end{align}
where $D(\alpha)$ is  the $D$-term, and 
\begin{align}
 [  {\cal B}(\beta,\alpha)]_+=     {\cal B}(\beta,\alpha) - \delta (\beta)  
  \int_{-1+|\alpha|} ^{1  -|\alpha|}    {\cal B} (\gamma,\alpha) \, d\gamma
\end{align}
is the  ``plus'' part of ${\cal B}(\beta,\alpha)$ that gives zero after integration 
over $\beta$.

\subsection{D term}

In principle, the function ${D(\alpha)}/{\alpha} $
may  have an unintegrable singularity for $\alpha=0$.
Then  it makes sense to split the D term contribution 
into the  ``plus part'' and $\delta (\alpha)$ contribution.
This will correspond to representation
\begin{align}
     {\cal B} (\beta,\alpha)=&  [  {\cal B}  (\beta,\alpha)]_+  +\delta (\beta) \left (\frac{D(\alpha)}{\alpha}  \right )_+
    \nonumber \\ & + \delta (\beta) \delta (\alpha) {\cal D} \ , 
      \label{B+D+}
\end{align}
with ${\cal D}$ corresponding to the ${\cal B}$ part of the matrix 
element of the local operator $\bar \psi (0) \gamma_\mu \psi (0)$.  

In fact, the   first  analysis  of the D term in the literature
\cite{Polyakov:1999gs}  uses  
the chiral soliton model expression for the pion DD, which in the present  notations may be written as 
\begin{align}
     {\cal B} (\beta,\alpha)= 
     \delta (\alpha) \frac{1}{|\beta|}   -\delta (\beta)  \frac{1}{|\alpha|}  \  , 
\end{align}
where  both the forward distribution $f(\beta)=
1/|\beta|$ and the D term part $D(\alpha)/\alpha = 
1/|\alpha|$ are singular. Note that we can rewrite 
this expression as
\begin{align}
     {\cal B} (\beta,\alpha)= 
     \delta (\alpha) \left (\frac{1}{|\beta|}  \right )_+  -\delta (\beta) \left (\frac{1}{|\alpha|}  \right )_+  \ ,
\end{align}
that  has the structure  of Eq. (\ref{B+D+})
 with the constant ${\cal D}$ equal to zero.

In what follows, we will assume that ${D(\alpha)}/{\alpha} $ is regular for $\alpha =0$,
so we will use the shorter representation (\ref{BplusD}).
In accordance with it,
 we  split GPD $B$ into the part coming from 
the ``plus'' part  of DD 
\begin{align}
 \frac{B_+(x,\xi) }{x}  \equiv & 
 \int_{\Omega}   {\cal B} (\beta,\alpha)   \,\biggl [   \delta (x - \beta -\xi \alpha) 
 - \delta (x - \xi \alpha )  \biggr ]
  \, d\beta \, d\alpha  
\label{GPDBplus}
\end{align}
and  that  generated by  the $D$-term
\begin{align}
 \frac{B_D(x,\xi) }{x}  \equiv & 
 \int_{-1} ^{1 }  \frac{D(\alpha)}{\alpha}   \delta (x - \xi \alpha) 
\, d\alpha  \  .
\label{GPDD}
\end{align}
The latter integral gives an  explicit expression 
\begin{align}
B_D(x,\xi) =  {\rm sign}(\xi)\,   \theta (|x|< |\xi|)\, D(x/\xi)\ ,
\label{Dexpl}
\end{align}
but sometimes it is instructive to  use the integral representation  as well.
In the $\xi=0$ limit, we obtain an 
 important relation
\begin{align}
 \frac{B_D(x,0) }{x}  = &  \delta (x) 
 \int_{-1} ^{1 }  \frac{D(\alpha)}{\alpha}  
\, d\alpha  \  . 
\label{GPDDxi0}
\end{align}
In other words, though the forward limit of $B(x,\xi)$ is $B(x,0)=-e(x)$,
the forward limit of $B(x,\xi)/x$ is 
\begin{align}
 \frac{B(x,0) }{x}  = -\left (\frac{e(x)}{x} \right )_+   + \delta (x) 
 \int_{-1} ^{1 }  \frac{D(\alpha)}{\alpha}  
\, d\alpha  \  . 
\label{GPDDxi0e}
\end{align}
A similar result holds for the border function:
\begin{align}
 \frac{B(x,x) }{x}  = -\left (\frac{E(x,x)}{x} \right )_+   + \delta (x) 
 \int_{-1} ^{1 }  \frac{D(\alpha)}{\alpha(1-\alpha)}  
\, d\alpha  \  . 
\label{GPDDxx}
\end{align}

\section{Sum rules for Nucleon GPDs} 

\subsection{Sum rule for $A(x,\xi)$}

Since $A(x, \xi)$ is given by  the simplest  DD representation
\begin{align}
 {A(x,\xi) }  =
 \int_{\Omega}    {\cal A}  (\beta,\alpha) \,  \delta (x - \beta -\xi \alpha)  \, d\beta \, d\alpha 
\label{GPDA2}
\end{align}
of Eq. (\ref{GPDF}) type, the sum rule 
 \begin{align}
  P\int_{-1}^1  \frac{A (x, x) - A(x,\xi)}{x-\xi}\,  dx  =  0  
  \label{eq_SRGA} 
 \end{align}
is  derived just as it was done for $F_{DD} (x,\xi)$ above (see Eqs. (\ref{eq_SRG2xx0}) --(\ref{HDDSR})).

\subsection{Sum rule   for $B(x,\xi)$}

\subsubsection{Naive construction}

Let us  now   apply the same construction for  the contribution 
 to   the sum rule 
(\ref{eq_SRG}) coming from  GPD
$B(x,\xi)$.
Using Eq.~(\ref{GPDB}), we have
\begin{align}  
P& \int_{-1}^1  \frac{B (x, x) }{x-\xi}\,  dx  
  \nonumber \\  = &
 P \int_{-1}^1   x \, \frac{dx}{x-\xi}\,    
 \int_{\Omega}   {\cal B} (\beta,\alpha)   \,
 \delta (x(1-\alpha) - \beta )  )    \, d\beta \, d\alpha   
     \\  &
     =   
P  \int_{\Omega}   {\cal B} (\beta,\alpha)   \,
 \,  
  \frac{\beta/ (1-\alpha)}{\beta- \xi (1-\alpha) }   \, d\beta \, d\alpha 
     \nonumber  \  , 
 \label{eq_SRG2xx} 
\end{align}
and
\begin{align}  
P&  \int_{-1}^1  \frac{ B(x,\xi)}{x-\xi}\,  dx  \\  & = 
  P \int_{-1}^1   x  \,  \frac{dx}{x-\xi}\,    
  \int_{\Omega}   {\cal B} (\beta,\alpha)   \,
\delta (x - \beta -\xi \alpha)  \, d\beta \, d\alpha      \nonumber
 \\ &
  =    
P \int_{\Omega}   {\cal B} (\beta,\alpha)   \,
   \, 
 \frac {\beta+ \xi \alpha  }{\beta- \xi (1-\alpha) }   \, d\beta \, d\alpha
     \nonumber  \\ & 
          =   
     P  \int_{\Omega}   {\cal B} (\beta,\alpha)   \,
         \, 
      \biggl [    \frac{\beta/ (1-\alpha)}{\beta- \xi (1-\alpha) } 
      - \frac{\alpha}{  1-\alpha }   \biggr ]  \, d\beta \, d\alpha
          \nonumber   \ . \nonumber
 \label{eqxxi} 
\end{align}
As a result,  
\begin{align}
 P\int_{-1}^1  \frac{B (x, x) - B(x,\xi)}{x-\xi}\,  dx   = 
  \int_{\Omega}       \frac{\alpha}{  1-\alpha }  \,    {\cal B} (\beta,\alpha)   \,
       \, d\beta \, d\alpha   
   \  .
\end{align}
Defining formally 
\begin{align}
   \int_{-1+|\alpha|} ^{1  -|\alpha|}        {\cal B} (\beta,\alpha)   \,
       \, d\beta =
    \frac{D(\alpha)}{\alpha}  
\end{align}
one obtains 
\begin{align}
 P\int_{-1}^1  \frac{B (x, x) - B(x,\xi)}{x-\xi}\,  dx   =    \int_{-1}^1  \frac{D(\alpha)}{1-\alpha} \, d \alpha 
   \  .
\end{align}
However, if the forward distribution $e(x)$ is singular for small $x$,
e.g., has a  Regge behavior $e(x) \sim x^{-a}$ with $a>0$,
we expect that ${\cal B} (\beta,\alpha)$ has a non-integrable 
singularity $\sim \beta^{-1-a}$ for $\beta \to 0$
(the function ${\cal B} (\beta,\alpha)$ is even in $\beta$ for the  $C$-even   ``quark + antiquark'' combination encountered 
in DVCS).

To avoid explicit infinities, we will     apply the ``DD$_+$+D'' separation   to   the generic relation 
(\ref{eq_SRG}). The derivation of the sum rule  then proceeds  in the following way.

\subsubsection{``Plus''  part}

Using Eq.~(\ref{GPDBplus}), we have
\begin{align}  &
P \int_{-1}^1  \frac{B_+ (x, x) }{x-\xi}\,  dx  
 = 
 P \int_{-1}^1   x \, \frac{dx}{x-\xi}\,    
 \int_{\Omega} {\cal B}  (\beta,\alpha)   \,
  \nonumber \\ & \times
  \biggl [ \delta (x(1-\alpha) - \beta ) -   \delta (x(1-\alpha) )  
     \biggr ]   \, d\beta \, d\alpha    \nonumber 
     \\ & 
     =   
P  \int_{\Omega}  {\cal B}  (\beta,\alpha)   \,
 \,  
  \biggl [ \frac{\beta/ (1-\alpha)}{\beta- \xi (1-\alpha) }  - 0
     \biggr ]    \, d\beta \, d\alpha  \  , 
 \label{eq_SRG2xx2} 
\end{align}
and
\begin{align}  &
P \int_{-1}^1  \frac{ B_+(x,\xi)}{x-\xi}\,  dx   = 
  P \int_{-1}^1   x  \,  \frac{dx}{x-\xi}\,    
  \int_{\Omega} {\cal B}  (\beta,\alpha)   \,
  \nonumber \\ & \times
  \biggl [ \delta (x - \beta -\xi \alpha)  
 - \delta (x - \xi \alpha )  \biggr ]   \, d\beta \, d\alpha    \nonumber 
 \\ &
  =    
P \int_{\Omega} {\cal B}   (\beta,\alpha)   \,
  \, 
  \biggl [ \frac {\beta+ \xi \alpha  }{\beta- \xi (1-\alpha) }+ \frac{\alpha}{  1-\alpha } 
     \biggr ]   \, d\beta \, d\alpha   \nonumber   \\ & 
          =   
     P  \int_{\Omega} {\cal B}  (\beta,\alpha)   \, 
        \frac{\beta/ (1-\alpha)}{\beta- \xi (1-\alpha) }             \, d\beta \, d\alpha  \ .
 \label{eqxxi3} 
\end{align}
Similarly to the  case of  GPD $A$, apparently  different delta-functions have converted
$1/(x-\xi)$  into identical  expressions.
As a result,  
\begin{align}
 P\int_{-1}^1  \frac{B_+ (x, x) }{x-\xi}\,  dx - P\int_{-1}^1  \frac{ B_+(x,\xi)}{x-\xi}\,  dx  =  0  \  .
\end{align}
Again,  we deal with the situation when the difference of  two integrals vanishes, 
but  each integral  does not necessarily vanish.

\subsubsection{``$D$''  part}

For the integral involving the border function, we have 
\begin{align}  &
 P \int_{-1}^1  \frac{B_D (x, x) }{x-\xi}\,  dx 
 =   0  \  , 
\label{HDxi}
   \end{align}
      because the  integrand  in (\ref{HDxi}) 
      vanishes for  \mbox{$x \neq 0$}  since   then  $B_{D}(x,x)=0$,
      while    for $x=0$ it is   given   by the $x \delta (x)$
distribution    that  produces zero  after  integration 
with a function that   is  finite   for $x=0$,  which  is  the case
 if  $\xi \neq 0$.     
    The second piece is given   by 
  \begin{align}  
P  \int_{-1}^1  \frac{ B_D(x,\xi)}{x-\xi}\,  dx  
  =  & 
  {\rm sgn}(\xi)\, P \int_{-1}^1  \frac{  \theta (|x|<|\xi|)  \,  D(x/\xi) }{x-\xi}\,  dx 
  \nonumber \\ & =  - \int_{-1}^1  \frac{D(\alpha)}{1-\alpha} \, d \alpha  \ .
 \label{SRD2ndB}
   \end{align}

For the difference of  the two  integrals we obtain
\begin{align} 
 & P \int_{-1}^1  \frac{B_D (x, x)}{x-\xi}  \,  dx  -
  P \int_{-1}^1  \frac{B_D(x,\xi)}{x-\xi}  \,  dx= \int_{-1}^1  \frac{D(\alpha)}{1-\alpha} \, d \alpha  \ . 
 \label{eq_SRG3} 
   \end{align}
%the same result as in Eq.~(\ref{eq_SRDD}).
Combining the results for the ``plus'' and $D$-parts gives 
\begin{align}
 P\int_{-1}^1  \frac{B (x, x) - B(x,\xi)}{x-\xi}\,  dx  = 
  \int_{-1}^1  \frac{D(\alpha)}{1-\alpha} \, d \alpha    \ ,
 \label{eq_SRBG} 
\end{align}
formally the same result as the naive construction.

Still, the importance of the ``DD$_+$+D'' separation should not be underestimated.
It emphasizes the fact that the D term cannot be defined 
simply as the result of integration of ${\cal B}   (\beta,\alpha)$ 
over $\beta$. The integral over $\beta$ diverges, and 
  the role of the D term in this  case 
   is to substitute   the divergent integral
   by a finite function $D(\alpha)/\alpha$.   In this sense, the ``DD$_+ + $D'' separation 
   serves as a renormalization prescription,  
and demonstrates the subtraction nature of the D term,
in particular, its role as a separate independent  entity that, in general, cannot be reconstructed 
from the behavior of the DD ${\cal B}   (\beta,\alpha)$  in the $\beta \neq 0$ region.

\subsection{Sum rules   for $H(x,\xi)$  and $E(x,\xi)$ }

Since $B(x,\xi) = - E(x,\xi)$, we have
\begin{align}
 P & \int_{-1}^1  \frac{E (x, \xi) }{x-\xi}\,  dx   \nonumber \\ & 
 =  P\int_{-1}^1  \frac{E (x, x) }{x-\xi}\,  dx -
  \int_{-1}^1  \frac{D(\alpha)}{1-\alpha} \, d \alpha    \ .
 \label{eq_SRE} 
\end{align}
Similarly, using $H(x,\xi)= A(x,\xi) + B(x,\xi) $, we obtain 
\begin{align}
 P & \int_{-1}^1  \frac{H (x, \xi) }{x-\xi}\,  dx   \nonumber \\ & 
 =  P\int_{-1}^1  \frac{H (x, x) }{x-\xi}\,  dx +
  \int_{-1}^1  \frac{D(\alpha)}{1-\alpha} \, d \alpha    \ .
 \label{eq_SRH} 
\end{align}

\section{Secondary GPD sum rule} 

\subsection{Compton form factors and border function}

  The derivation above confirms the result that the   Compton form factor  ${\cal C}_H (\xi)$  is expressed 
    through a one-dimensional {\it border function} $H(\xi, \xi) \equiv {\mathfrak h} (\xi)$
     \begin{align}
{\cal C} _H (\xi) = 
-i \pi {\mathfrak h} (\xi) + 
 P\int_{-1}^1  \frac{{\mathfrak h} (x)}{x-\xi}\,  dx + \Delta \ .
  \label{eq_Cxi2} 
 \end{align}
  The integral over $x$ here has opposite infinitely large  canceling 
 contributions from $x \sim \xi$ region where ${\mathfrak h} (x) \sim {\mathfrak h} (\xi)$.
  One can exclude   them by writing
     \begin{align}
{\cal C}_H (\xi) = &
-i \pi {\mathfrak h} (\xi) + 
 P\int_{-1}^1  \frac{{\mathfrak h} (\xi)}{x-\xi}\,  dx  \nonumber \\ &+ 
P  \int_{-1}^1  \frac{{\mathfrak h} (x)-{\mathfrak h} (\xi)}{x-\xi}\,  dx+ \Delta \ .
  \label{eq_Cxi22} 
 \end{align}
 The first integral here can be taken explicitly, and the second one 
 is regular for $x=\xi$ needing thus no $P$ prescription.
 As a result,
      \begin{align}
{\cal C}_H (\xi) = &
 {\mathfrak h} (\xi) [ \ln (1- x_{\rm Bj})-i \pi]   \nonumber \\ &
 + 
 \int_{-1}^1  \frac{{\mathfrak h} (x)-{\mathfrak h} (\xi)}{x-\xi}\,  dx+ \Delta \ ,
  \label{eq_Cxi3} 
 \end{align}
 where $x_{\rm Bj} = 2\xi/(1+\xi)$ is the Bjorken variable.

\subsection{Proposing secondary sum rule}

  In this formulation, the description of DVCS becomes similar 
    to that of DIS, with the  change of usual parton distributions 
    $f(x)$ by border functions ${\mathfrak h} (x)$. 
    So,  what can we say about 
     the difference between these two functions?
    
  When the D term $D(x/\xi)$ vanishes at the border point  $x=\xi$
 (which is the usual assumption), it is not visible
 in the border function ${\mathfrak h} (x)$. 
 It is also not visible in the forward distribution  $f(x)$,
 i.e., these two functions are apparently determined by the DD part
 of GPD $H(x,\xi)$   only. 
Now, if  one takes formally   the  $\xi \to 0$ limit of the sum rule 
(\ref{HDDSR})  for the DD part,  one  arrives at the 
sum  rule 
\begin{align}
 \int_{-1}^1  \frac{H_{DD} (x, x) - H_{DD} (x,0)}{x}\,  dx  \stackrel{?}{=}  0
 \label{eq_SR} 
\end{align}
or
\begin{align}
 \int_{-1}^1  \frac{{\mathfrak h}  (x) - f(x)}{x}\,  dx  \stackrel{?}{=}    0
 \label{eq_SRfb} 
\end{align}
relating the ``minus first moments'' of these two functions.

Note that 
for the $C$-even amplitudes  studied in DVCS and DIS,    both
$f(x)/x$  and ${\mathfrak h}(x)/x$ are even functions
of $x$.  Hence, one may rewrite  Eq. (\ref{eq_SRfb})
as 
\begin{align}
 \int_{0}^1  \frac{{\mathfrak h}  (x) - f(x)}{x}\,  dx  \stackrel{?}{=}    0 \ .
 \label{eq_SRfb2} 
\end{align}
It is straightforward to see that this relation cannot be true as a  general statement.
E.g., take a constant DD,   
$${\cal F} ^{\rm const\, DD} (\beta, \alpha)=
\frac12 {\rm sgn} (\beta) 
\theta (|\beta| +|\alpha| \leq 1)\ . $$
Then $$f(x)^{\rm const\, DD} =  {\rm sgn}  (x) (1-|x|)\, ,$$  while 
$${\mathfrak h} ^{\rm const\, DD}  (x) =\frac{ {\rm sgn}  (x) }{1+|x|} \  .$$
Thus, $f(x)^{\rm const\, DD} < {\mathfrak h} ^{\rm const\, DD}  (x)$ for $x>0$. 
In general,  ${\mathfrak h}  (x)  > f(x)$  for positive $x$ if one uses the 
usual modeling of GPDs based on factorized DD Ansatz 
with positively definite,  monotonically decreasing $f(x)$. 
Thus, in all these cases,  the sum rule (\ref{eq_SRfb2}) does not hold.
Moreover,  if the difference ${\mathfrak h}  (x) - f(x)$ 
does not vanish in the $x \to 0$ limit, then the integral in 
(\ref{eq_SRfb2})  simply diverges. 
Since none of the functions ${\mathfrak h}  (x)$, $ f(x)$ 
is expected to vanish for $x=0$, the integrals 
for ${\mathfrak h}  (x) /x$ and 
$ f(x)/{x}$  separately diverge. 
Obviously,   singularities of even functions  ${\mathfrak h}  (x) /x$ and 
$ f(x)/{x}$  for $x=0$
cannot  be regularized by the principle value prescription.

Furthermore, if one takes formally $\xi =0$ in the basic sum rule 
(\ref{eq_SRH})  one gets
 \begin{align}
 P \int_{-1}^1  \frac{H  (x, x) - H (x,0)}{x}\,  dx  \stackrel{?}{=}  \Delta  \  .
 \label{eq_SR0} 
\end{align}
Again, the integrand involves an even function of $x$
with a non-integrable singularity for $x=0$.

\subsubsection{Secondary sum rule for GPD $B(x, \xi)$}

A puzzling feature  of Eq. (\ref{eq_SR0}) is the  presence of the D-term-dependent
constant $\Delta$ for a sum rule involving functions 
$H  (x, x)$ and $ H (x,0)$ that apparently are insensitive to 
the D term contribution.
However, as we have seen in \mbox{Sec. III C,} the ``DD$_+$+D'' separation gives
a $\delta (x)$ contribution for  $ B  (x, x) /x$  and  $B (x,0)/x$
with coefficients proportional to integrals of the D term.
Through $H=A+B$ these contributions appear in expressions for GPD  $H$  as well.
In particular, one  can easily  see from Eq. (\ref{GPDDxi0})
 that
\begin{align}
 \int_{-1}^1  \frac{B(x,\xi) }{x}  \,  dx  = \int_{-1} ^{1 }  \frac{D(\alpha)}{\alpha}  
  \label{eq_SR+} 
\end{align}
for any $\xi$, including $\xi=0$. In the $\xi$=0 limit this gives 
\begin{align}
 \int_{-1}^1  \frac{B(x,0) }{x}  \,  dx  = \int_{-1} ^{1 }  \frac{D(\alpha)}{\alpha}  \  . 
  \label{eq_SR+0} 
\end{align}

For the integral involving the border function, using Eq. (\ref{GPDDxx}) we get
\begin{align}
 \int_{-1}^1  \frac{B(x,x) }{x}  \, & dx  =
  \int_{-1} ^{1 }  \frac{D(\alpha)}{\alpha(1-\alpha)}  
\, d\alpha   \  .
\end{align}
Thus,
\begin{align}
 \int_{-1}^1  \frac{B(x,x) -B(x,0)}{x}  \, & dx  =
  \int_{-1} ^{1 }  \frac{D(\alpha)}{1-\alpha}  
\, d\alpha \equiv \Delta   \  .
\label{BSR2nd}
\end{align}

The nonzero result on the right hand side comes 
from the $\delta (x)$ terms in the expressions (\ref{GPDDxi0}),(\ref{GPDDxx}) for $B(x,x)/x$
and $B(x,0)/x$.  The remaining terms $[B(x,x)/x]_+$ and $[B(x,0)/x]_+$
have ``plus''-structure and  each automatically produces zero contribution 
into the sum rule. In other words, the actual difference 
between the border function $B(x,x)$ and the forward distribution 
$B(x,0)$ in the region $x \neq 0$ has no reflection in the sum rule
(\ref{BSR2nd}).

\subsubsection{Secondary sum rule for GPD $A(x, \xi)$}

Similarly, taking formally the $\xi \to 0$ limit of the  sum rule 
 \begin{align}
  P\int_{-1}^1  \frac{A (x, x) - A(x,\xi)}{x-\xi}\,  dx  =  0  
  \label{eq_SRGA2} 
 \end{align}
 for GPD $A(x,\xi)$, one deals with the difference of two 
 divergent integrals
  \begin{align}
  P\int_{-1}^1  \frac{A (x, x) - A(x,0)}{x}\,  dx   \stackrel{?}{=}    0  \ ,
  \label{eq_SRGA2?} 
 \end{align}
which cannot be regularized by the principal value prescription.
Since there is no D term for $A(x,\xi)$,  
the idea of ``DD$_+$+D'' separation only suggests to treat 
$A (x, x)/x$ and  $A(x,0)/x$ as ``plus'' distributions,
in which case each term in (\ref{eq_SRGA2?})  gives 
an automatic zero.
Again, the resulting ``sum rule'' contains no information about
the difference between $A (x, x)$ and  $A(x,0)$.  

Summarizing, the $\xi \to 0$ limit of the dispersion sum rule
(\ref{eq_SRG}) produces divergent integrals, which may be 
regularized by using ``DD$_+$+D'' separation,
but the resulting sum rule contains  no information about 
the difference between  border function and  forward distribution. 

However, one may try to study this difference
incorporating various models for GPDs,
in particular, the {\it factorized DD Ansatz}.

 \section{Modeling border functions} 
   
 \subsection{Border  function  $A(x,x)$} 

 \subsubsection{Factorized DD Ansatz}

In the forward limit  $\xi=0$, GPD  $A(x,\xi)$ converts into the forward 
distribution $f(x)+e(x)\equiv a(x) $, i.e., 
\begin{align}
a(x) &=
    \int_{-1+|x|} ^{1  -|x|}  {\cal A} (x,\alpha) \,   d\alpha   \   .
\label{GPDtoax}
\end{align}
Thus,  the forward  distribution $a(x)$  is  obtained by integrating 
over vertical lines $\beta =x$  in the $(\beta,\alpha)$ plane.
For nonzero $\xi$, GPDs  are obtained from  DDs  through  integrating them 
along   the lines $\beta=x-\xi \alpha$ having   $1/\xi$ slope. 
The reduction formula
  (\ref{GPDtoax}) 
suggests   the {\it factorized DD Ansatz} 
\begin{align}
{\cal A} (\beta,\alpha) =  h_A(\beta,\alpha) \, {a(\beta)} \   ,
\label{FDDA}
\end{align}
where $a(\beta)$   is the forward  distribution, while  $h_A(\beta,\alpha) $  
determines DD profile in the 
$\alpha$  direction  and satisfies the normalization condition
\begin{align}
 \int_{-1+|\beta |} ^{1  -|\beta|}     h_A(\beta,\alpha) \,   d\alpha =1\   .
\label{hnorm}
\end{align}
The usual choice for the profile 
 function  is
 \begin{align}
 h^{(N)}(\beta,\alpha) = \frac{\Gamma (2N+2)}{2^{2N+1} \Gamma^2 (N+1)}
\frac{[(1- |\beta|)^2 - \alpha^2]^N}{(1-|\beta|)^{2N+1}} \,  .  \label{modn} 
 \end{align}
The  width of the profile  is governed by the parameter $N$.

The border function corresponding to such an Ansatz
for positive $x$  is given by
 \begin{align}
 A^{(N)}(x,x)=& 
\frac{(1- x)^N }{(1+x)^{N+1}}\frac{\Gamma (2N+2)}{ \Gamma^2 (N+1)}
 \nonumber \\ & \times 
\int_0^1\gamma^N (1-\gamma)^N  \varphi^{(N)} (x_0 \gamma) 
 \, 
 d\gamma
 \,  ,  \label{AxxN} 
 \end{align}
where $x_0=2x/(1+x)$ and
 \begin{align}
 \varphi^{(N)} (\beta) = \frac{a(\beta)}{(1-\beta)^{2N+1}} \ . 
 \end{align}
%Recall that for $C$-even GPDs we have $A(-x,x)=-A(x,x)$.

 \subsubsection{Border function  $H_A (x,x)$}

         Since $A=H+E$, with the forward limits $f(x)$ and $e(x)$ of the
         functions $H (x,\xi)$ and $E(x,\xi)$ having, in general,
         different $x$-dependence, it makes sense to use different
         factorized {\" A}nsatze for these two parts.
         It is usually assumed that $f(x)$ and $e(x)$ have the same $x^{-a}$ 
         Regge behavior for small $x$, but differ in the $x \to 1$ region,
         with $f(x)$ having $\sim (1-x)^3$  behavior and $e(x)$ being closer to
         $\sim (1-x)^5$.  
 
As an example, let us  take the forward function 
\mbox{$f(\beta) = \beta^{-a} (1- \beta)^3$} 
with $\beta^{-a}$ Regge behavior for small $\beta$ and the 
usual \mbox{$(1-\beta)^3$}  behavior for $\beta \to 1$. 
Choosing  $N=1$ profile, we have
 \begin{align}
 H_A^{(1)}(x,x)=& 
\frac{1- x }{(1+x)^{2}}% \nonumber \\ & \times 
  \frac{x_0^{-a}} {(1-a/2)(1-a/3)} \ . 
 \end{align}
 Changing to $N=2$ profile produces
  \begin{align}
 &H_A^{(2)}(x,x)= 
\frac{1- x }{(1+x)^{2}} 
  \frac{x_0^{-a}} {(1-a/3)(1-a/4)(1-a/5)} 
\\ & \times 
  \left [5-a- (4-a -(2-a)x_0 ){_2 F_1 (1,3-a,6-a,x_0)}   \right ]\ . 
     \nonumber 
 \end{align}
 
  Note that imaginary part of the Compton amplitude 
      is given by border function for $x=\xi = x_{\rm Bj}/(2- x_{\rm Bj})$.
      In this case, $x_0 \to x_{\rm Bj}$ and   $(1-x)/(1+x) \to 1- x_{\rm Bj}$.
      In these variables,
        \begin{align}
       H_A^{(N)}(\xi,\xi)=& \left (1- \frac{x_{\rm Bj}}{2} \right )
(1- x_{\rm Bj})^N \frac{\Gamma (2N+2)}{ \Gamma^2 (N+1)}
 \nonumber \\ & \times 
\int_0^1\gamma^N (1-\gamma)^N  \varphi^{(N)} ( x_{\rm Bj} \gamma) 
 \, 
 d\gamma
 \,  .  \label{AxxNB} 
 \end{align}   
 For small $x_{\rm Bj}$, we have 
  \begin{align}
 H_A^{(1)}(\xi,\xi)=&  
  \frac{x_{\rm Bj}^{-a}} {(1-a/2)(1-a/3)} +\ldots  
 \end{align}
      and 
        \begin{align}
 &H_A^{(2)}(\xi,\xi)= 
  \frac{x_{\rm Bj}^{-a}} {(1-a/3)(1-a/4)(1-a/5)}   +\ldots   \ ,
 \end{align}
 i.e., the same     (up to overall factor) Regge    behavior  
      $ x_{\rm Bj}^{-a}$ as in case of forward distribution.
      For $a=1/2$, the {\it GPD enhancement factor} 
     (see, e.g., Ref. \cite{Radyushkin:1998es})
  \begin{align}
  \left.       R \equiv \frac{H_A(\xi, \xi)}{f(x_{\rm Bj})} \right |_{x_{\rm Bj} \to 0}
      \end{align}   
      equals 8/5=1.6  in case of $N=1$ profile, and 32/21$\approx 1.5$ for $N=2$.
      Taking the $N\to \infty$ limit, we obtain an infinitely narrow profile,
      and $H_A(\xi,\xi)$ coincides with $f (\xi)$.
      Then $R\to (\xi/x_{\rm Bj})^{-a} \to 2^a$, which is $\approx 1.4$ for $a=0.5$.
      Thus, $R$ does not change significantly when the profile broadens 
      from the  infinitely narrow  one  to that corresponding to $N=1$.
      
      For a flat $N=0$ profile, we have 
       \begin{align}
 H_A^{(0)}(x,x)=& 
\frac{x_0^{-a}}{1+x}% \nonumber \\ & \times 
 \left [ \frac1{1 - a} -\frac{ x_0}{1-a/2}  + \frac{x_0^2}{3 - a} \right ] 
 \end{align}
      and $R=1/(1-a)$, i.e., $R=2$ for $a=1/2$.

         \subsubsection{Border function  $E_A (x,x)$}

         For modeling  $E_A(x,\xi )$, we will  take $e(x)=x^{-a} (1-x)^5 $ 
         as the forward limit. In this case, the simplest analytic expression 
         is obtained for $N=2$ profile, which gives
        \begin{align}
 E_A^{(2)}(x,x)=& 
\frac{(1- x)^2 }{(1+x)^{3}}% \nonumber \\ & \times 
  \frac{x_0^{-a}} {(1-a/3)(1-a/4)(1-a/5)} 
 \end{align}
         or
                \begin{align}
 &E_A^{(2)}(\xi,\xi)= 
  \frac{x_{\rm Bj}^{-a} (1- x_{\rm Bj})^2
   \left (1- {x_{\rm Bj}}/{2} \right )} {(1-a/3)(1-a/4)(1-a/5)}    \ .
 \end{align}
        For $N=1$ profile, the result is also rather simple:
           \begin{align}
 E_A^{(1)}(x,x)&= 
\frac{1- x }{(1+x)^{2}} 
x_0^{-a}   \left [ \frac1{(1 - a/2)(1-a/3)} 
\right.  \\   - & \left. \frac{ x_0}{(1-a/3)(1-a/4)}  + 
\frac{6 x_0^2}{(4-a)(5-a)} \right ]  \  .  \nonumber
 \end{align}
        
         \subsection{Border  function  $B(x,x)$} 
         
  It should be emphasized that the model 
  $A(x,\xi)=H_A(x,\xi) +  E_A(x,\xi)$  is just a model for the sum 
         $H(x,\xi) + E(x,\xi)$, with $H(x,\xi)$ and $  E(x,\xi)$
         not necessarily coinciding with $H_A(x,\xi)$ and $  E_A(x,\xi)$.
         This situation is similar to the ``DD plus D'' scenario,
         where  one has
        $H(x,\xi) = H_{DD} (x,\xi) +D$ term and 
       $E(x,\xi) = E_{DD} (x,\xi) -D$ term, so that 
        $H_{DD} (x,\xi) + E_{DD} (x,\xi) $ gives a model for 
        $H(x,\xi) + E(x,\xi)$  while   $H(x,\xi) \neq
        H_{DD} (x,\xi) $ and  $ E(x,\xi)  \neq
        E_{DD} (x,\xi)$.   As shown in Ref.   \cite{Radyushkin:2013hca}, 
        the difference between $H(x,\xi)$ and 
        $H_A(x,\xi)$ (and between $  E(x,\xi)$ and $E_A(x,\xi)$)   
              in our construction is even more  serious: it 
         does not reduce to the D term only and contains a term that changes the 
         border function. 
         
         The  strategy described in our paper 
         \cite{Radyushkin:2013hca}  is to build a model for $B(x,\xi)=-E(x,\xi)$,
       using the DD representation (\ref{GPDB})
         and then get a model for $H(x,\xi)= A(x,\xi)+B(x,\xi)$.

    \subsubsection{Structure of  $B(x,\xi)$} 
    
    Taking the forward limit $\xi=0$ in Eq. (\ref{GPDB}) 
    that defines  $B(x,\xi)$, we get
   \begin{align}
b(x) &= -e(x) = 
  x  \int_{-1+|x|} ^{1  -|x|}  {\cal B} (x,\alpha) \,   d\alpha  \  . \label{GPDtobx} 
  \end{align} 
  This reduction formula
suggests   the Ansatz 
\begin{align}
{\cal B} (\beta,\alpha) =  -  h_B (\beta,\alpha) \frac{e  (\beta)}{\beta} \   ,
\label{FDDABh}
\end{align}
that reconstructs DD  ${\cal B} (\beta,\alpha)$ from the forward 
function $e (\beta)/{\beta}$ that has an extra factor $1/\beta$ singular 
in the $\beta \to 0$ limit. 
In general, we can define 
\begin{align}
{\cal B} (\beta,\alpha) =  -  \frac{{\cal E}_B (\beta,\alpha) }{\beta} \   ,
\label{FDDAB}
\end{align}
with DD ${\cal E}_B (\beta,\alpha)$  having the same projection 
\begin{align}
e (\beta) &=
    \int_{-1+|\beta|} ^{1  -|\beta|}  {\cal E}_B (\beta,\alpha) \,   d\alpha   \ ,
\label{GPDtoabeta}
\end{align}
on the $\beta$ axis as ${\cal E}_A (\beta,\alpha)$.
Because of possible singularity of  ${\cal E}_B (\beta,\alpha)/\beta$  at $\beta=0$,
we write it in the ``DD$_++ D$''  representation:
\begin{align}
{\cal B} (\beta,\alpha) = - \left ( \frac{{\cal E}_B (\beta,\alpha)}{\beta}  \right )_+  
+
\delta (\beta) \frac{D(\alpha) }{\alpha}  \  ,
\end{align}
where $D(\alpha)$ is  the $D$-term.
As a result, we have
\begin{align}
 {B(x,\xi) }  =  &
 -  
  x \int_{\Omega}   \left [ \left ( \frac{{\cal E}_B (\beta,\alpha)}{\beta}  \right )_+ -
  \delta (\beta) \frac{D(\alpha) }{\alpha} \right ]  \nonumber \\ &
  \times  \delta (x - \beta -\xi \alpha)  \, d\beta \, d\alpha \nonumber \\ &
  = -  E_{B+}(x,\xi)  +{\rm sgn} (\xi) \,    D(x/\xi)  \ , \nonumber 
  \end{align}
  where
  \begin{align}
  \frac{E_{B+} (x,\xi)}{  x } =& \int_{\Omega}  \left( \frac{{\cal E}_B(\beta,\alpha)}{\beta}  \right )_+    
          \delta (x - \beta -\xi \alpha)   \, d\beta \, d\alpha       \nonumber \\ & = 
            \int_{\Omega}  \frac{{\cal E}_B(\beta,\alpha)}{\beta}     \Big [ 
    \,  \delta (x - \beta -\xi \alpha) \nonumber  \\ &   - \delta (x -\xi \alpha)      \Big ]  \, d\beta \, d\alpha    \  .
      \label{GPDE+}
\end{align}
%has the  structure  of  a one-DD representation.  
Since 
$E_{B+} (x,\xi)/  x $ is built from the 
``plus'' part of a DD it should satisfy 
\begin{align}
\int_{-1}^1 \, E_{B+} (x,\xi) \, dx = \int_{\Omega}  
                           \left [  \frac{{\cal E}_B (\beta,\alpha)}{\beta}  \right ]_+ \, 
                                    \, d\beta \, d\alpha = 0 \ 
 . 
\label{H+zero}
\end{align}
 Being (for $C$-even combination) an even function of $x$,  the function $E_{B+} (x,\xi)/x$ obeys 
\begin{align}
\int_0^1   \frac{E_{B+} (x,\xi)}{  x } \, dx =0 \ .
\end{align}

Using the relation $x=\beta + \xi \alpha$, we may extract from 
$E_{B+} (x,\xi)$  the component $E_{B} (x,\xi)$  that 
is obtained from DD ${\cal E}_B (\beta,\alpha)$ not divided by $\beta$.
Namely, the function $E_{B+} (x,\xi)$ may be displayed as 
%\begin{widetext} 
 \begin{align}
&  E_{B+}  (x,\xi) = x \int_{\Omega}  
   \frac{ {\cal E}_B (\beta,\alpha)}{\beta} \left [ 
    \,  \delta (x - \beta -\xi \alpha)  \right. 
    \nonumber \\ & \left.  \hspace{4cm} -  \delta (x  -\xi \alpha) 
     \right ] \, d\beta \, d\alpha
     \nonumber \\ & \hspace{2cm}
       =\int_{\Omega}  {\cal E}_B (\beta,\alpha)
           \,  \delta (x - \beta -\xi \alpha)    \, d\beta \, d\alpha
       \nonumber \\ &    
       +\xi  \int_{\Omega}  
           \frac{\alpha}{\beta}\,  {\cal E}_B (\beta,\alpha) \, 
           \left [  
             \delta (x - \beta -\xi \alpha)  -   \delta (x  -\xi \alpha) 
            \right ] \, d\beta \, d\alpha
              \nonumber \\ & \hspace{2cm}
                   \equiv E_{B} (x,\xi)  +{\xi } E^1_{B+} (x,\xi)
 \  ,
\label{GPDE1}
\end{align}
where
\begin{align}
E_{B} (x,\xi) \equiv   \int_{\Omega}  
                     {\cal E}_B (\beta,\alpha)  \, 
                                     \delta (x - \beta -\xi \alpha) \, d\beta \, d\alpha 
 \end{align}
 is constructed from  ${\cal E}_B (\beta,\alpha)$ in the same way as 
$E_{A} (x,\xi)  $ is obtained from ${\cal E}_A (\beta,\alpha)$, and  
\begin{align}
{E^1_{B+} (x,\xi)} \equiv   \int_{\Omega}  
                        \left ( \frac{\alpha}{\beta}\,  {\cal E}_B (\beta,\alpha) \right )_+ \, 
                                     \delta (x - \beta -\xi \alpha) \, d\beta \, d\alpha
 \end{align}
 is the extra term. 
Since $E^1_{B+} (x,\xi) $ is built from
the ``plus'' part of a DD, its $x$-integral
from $-1$ to 1 is equal to zero, but 
in fact it vanishes also for a simpler reason
that $E^1_{B+} (x,\xi) $  is an odd function of $x$.
So, in this case, we cannot make any conclusions about the  magnitude of the 
$x$-integral of $E^1_{B+} (x,\xi) $ 
from  0 to 1. 

Thus,   we can  represent GPD $B$ as 
\begin{align}
 {B(x,\xi) }  =  &
  - E_{B} (x,\xi)  - {\xi } E^1_{B+} (x,\xi)+{\rm sgn} (\xi) \,    D(x/\xi)  \ ,
  \end{align}
and GPD $H$ as 
\begin{align}
 {H(x,\xi) }  =  & H_{A} (x,\xi)  +E_{A} (x,\xi) 
  - E_{B} (x,\xi)  \nonumber \\ & - {\xi } E^1_{B+} (x,\xi)+{\rm sgn} (\xi) \,    D(x/\xi)  \  .
  \end{align}
This formula is quite general.  In particular, it does not involve 
factorized DD Ansatz assumptions.

The simplest model assumption 
is  ${\cal E}_A (\beta,\alpha)={\cal E}_B (\beta,\alpha)$.
It was used 
in our paper \cite{Radyushkin:2013hca}.
However, if one chooses different profile functions  $h_i (\beta, \alpha)$
when representing  ${\cal E}_i (\beta,\alpha) = h_i (\beta, \alpha) e(\beta)$
(with $i=A,B$) one would get, in general, different results 
for   $E_{A} (x,\xi) $ and 
  $ E_{B} (x,\xi) $. 
  The modeling of ${\cal E}_i (\beta,\alpha) $ is performed in the same way as 
  for $H_A (x,\xi)$.  The new element is modeling of  $E^1_{B+} (x,\xi)$.
However, in practice it is simpler 
to build a model for the whole function $E_{B+} (x,\xi)$,
and build GPD $H$ using 
\begin{align}
 {H(x,\xi) }  =  & H_{A} (x,\xi)  +E_{A} (x,\xi) 
  - E_{B+} (x,\xi)    \nonumber \\ & +{\rm sgn} (\xi) \,    D(x/\xi)  \  .
  \end{align}

       \subsubsection{Border  function  $E_{B+} (x,x)$} 
       
   We can get a factorized DD model for the border 
   function  $E_{B+} (x,x)$ by ``recycling'' our results 
   for $E_{A} (x,x)$:  we should just change $a \to a+1$
   and add the $x$ factor    in the examples considered above.
   In particular, with $N=2$ profile, we get 
            \begin{align}
 E_{B+}^{(2)}(x,x)=& 
 \frac{5}{4} \, 
\left ( \frac{1- x }{1+x} \right )^2 % \nonumber \\ & \times 
  \frac{x_0^{-a}} {(1-a/2)(1-a/3)(1-a/4)} 
 \end{align}
         or
                \begin{align}
 &E_{B+}^{(2)}(\xi,\xi)=  \frac{5}{4}
  \frac{x_{\rm Bj}^{-a} (1- x_{\rm Bj})^2
} {(1-a/2)(1-a/3)(1-a/4)}    \ .
 \end{align}
 As expected,  $E_{B+}^{(2)}(\xi,\xi) $ is larger than $E_{A}^{(2)}(\xi,\xi) $:
    \begin{align}
 &\frac{E_{B+}^{(2)}(\xi,\xi)}{E_{A}^{(2)}(\xi,\xi) } =  \frac{5-a}{(2-a)
(2- x_{\rm Bj}) } 
  \ , 
 \end{align}
 and, thus, the difference $E_{A} (\xi,\xi) 
  - E_{B+} (\xi ,\xi)  $ is negative
  (for positive $\xi$).

      For $N=1$ profile, we have 
           \begin{align}
 E_{B+}^{(1)}(x,x)&= 
\frac{1- x }{1+x} 
x_0^{-a}   \left [ \frac{3/2}{(1 - a)(1-a/2)} 
\right.  \\   - & \left. \frac{ x_0}{(1-a/2)(1-a/3)}  + 
\frac{ x_0^2}{4 (1-a/3)(1-a/4)} \right ]  \  .  \nonumber
 \end{align}    
 Again,  $E_{B+}^{(1)}(\xi,\xi) > E_{A}^{(1)}(\xi,\xi) $.

   \subsubsection{Two-DD representation of the ${\cal B} $ part}  
    
   In the definition (\ref{pmurmu2})   of the  nucleon DDs, the ${\cal B}$ 
   DD was accompanied by the $2\beta (Pz) +\alpha (rz) $
   factor, which corresponds to ``one-DD'' representation.
   In principle, one can also use the ``two-DD'' representation,
   in which this contribution is given by expression 
      \begin{align}
      2 (Pz) {\cal P} (\beta,\alpha) + (rz)  {\cal R} (\beta,\alpha)   
      \end{align}  
   involving two DDs, ${\cal P}$ and ${\cal R}$. 
   The two-DD representation is redundant, in the sense that the 
   ``gauge transformation''
    \begin{align}
 {\cal P} (\beta, \alpha) &\to  {\cal P}  (\beta, \alpha) +\partial \chi (\beta,\alpha)/\partial \alpha  \ , 
 \label{Fgauge} \\
 {\cal R} (\beta, \alpha) &\to  {\cal R} (\beta, \alpha) -\partial \chi (\beta,\alpha)/\partial \beta  
\label{Ggauge} 
\end{align} 
   does not change the total contribution. The one-DD representation corresponds
   to  the gauge in which $ {\cal P} (\beta, \alpha) = \beta  {\cal B} (\beta, \alpha)$
   and  ${\cal R} (\beta, \alpha) = \alpha  {\cal B} (\beta, \alpha)$.
   The D term contribution is contained in the ${\cal R} (\beta, \alpha) $ DD, 
   \begin{align}
  {D(\alpha)}   =  \int_{-1+|\alpha|} ^{1  -|\alpha|}        {\cal R} (\beta,\alpha)   \,
       \, d\beta  \ ,
\end{align}
    and it cannot 
   be changed by a gauge transformation.  However, the remaining terms in ${\cal R}$
 may be totally  reshuffled 
   into ${\cal P}(\beta,\alpha)$ using the gauge function
  \begin{align} 
\chi_{D} (\beta,\alpha) =&-{{\rm sgn}(\beta)} 
%\int^{-|\beta|}_{-1+|\alpha|}{\cal R} (\gamma,\alpha) \, d \gamma  %\right. \nonumber \\ & \left. +
\int_{ |\beta| }^{1-|\alpha|} {\cal R} (\gamma,\alpha) \, d \gamma %\right \} 
\label{D_gauge}
 \end{align}  
(cf. \cite{Teryaev:2001qm,Tiburzi:2004qr}). 
As a result,  $ 2 (Pz) {\cal P} (\beta,\alpha) + (rz)  {\cal R} (\beta,\alpha)    $
converts into the expression 
 \begin{align} 
 2 (Pz) {\cal P}_{D}(\beta, \alpha)  +  (rz) \delta (\beta) D(\alpha)
 \label{PDplusD}
 \end{align}  
 in which the  ${\cal R}_D (\beta,\alpha) $ DD reduces to the D term.
 Using these relations, one can find connection between
$ {\cal P}_{D}(\beta, \alpha) $ and  the one-DD function ${\cal B}$.
Writing ${\cal B} (\beta, \alpha)$ as $- e(\beta)h_B (\beta,\alpha)/\beta$,
we obtain 
   \begin{align}
{\cal P}_{D} (\beta,\alpha) = & - e(\beta)h_B (\beta,\alpha)  \label{PD}
  \\ & 
+ {{\rm sgn}(\beta)}
\int_{ |\beta| }^{1-|\alpha|} \frac{e(\gamma)}{\gamma} 
 \frac{\partial}{\partial \alpha} \Big [ \alpha h_B (\gamma,\alpha) \Big  ]  \, d \gamma 
  \  .   \nonumber
  \end{align}  
 
   Since the total sum of terms related to $(Pz)$ and $(rz)$ structures 
   is not changed under the gauge transformation,
   the GPD $P_D (x,\xi)$ obtained in ``$D$-gauge'' should  coincide 
   with GPD $B_+ (x,\xi)=-E_{B+} (x, \xi)$ obtained in the ``one-DD'' gauge. 
   In this sense, the expression for ${\cal P}_{D} (\beta,\alpha)$ given above corresponds to 
   the decomposition $E_{B+} (x,\xi) = E_{B} (x,\xi)  +{\xi } E^1_{B+} (x,\xi)$ 
   of Eq. (\ref{GPDE1}).

\begin{figure}[h]
 \begin{center} 
 \includegraphics[scale=0.21]{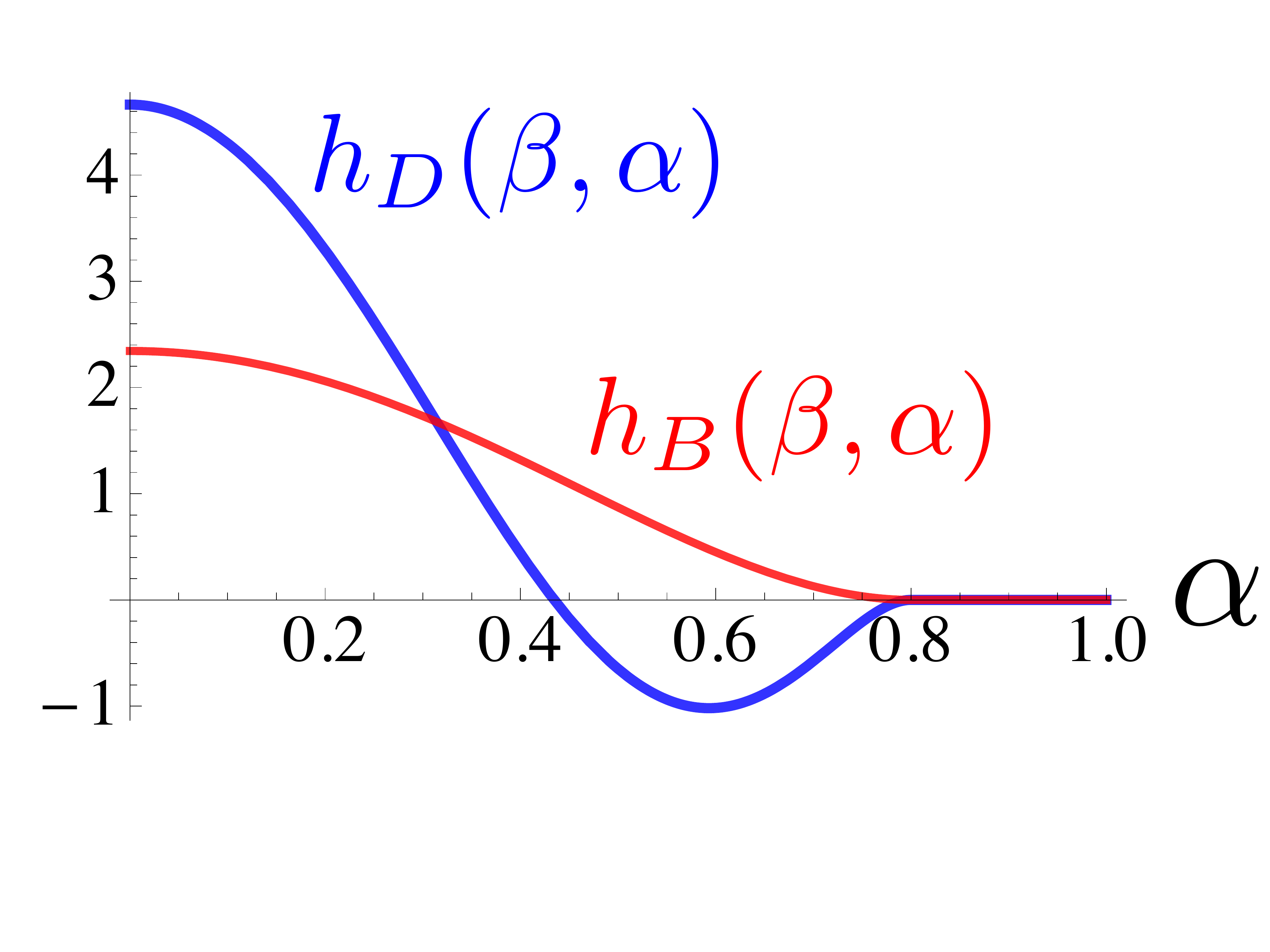} \hspace{2cm} \includegraphics[scale=0.2]{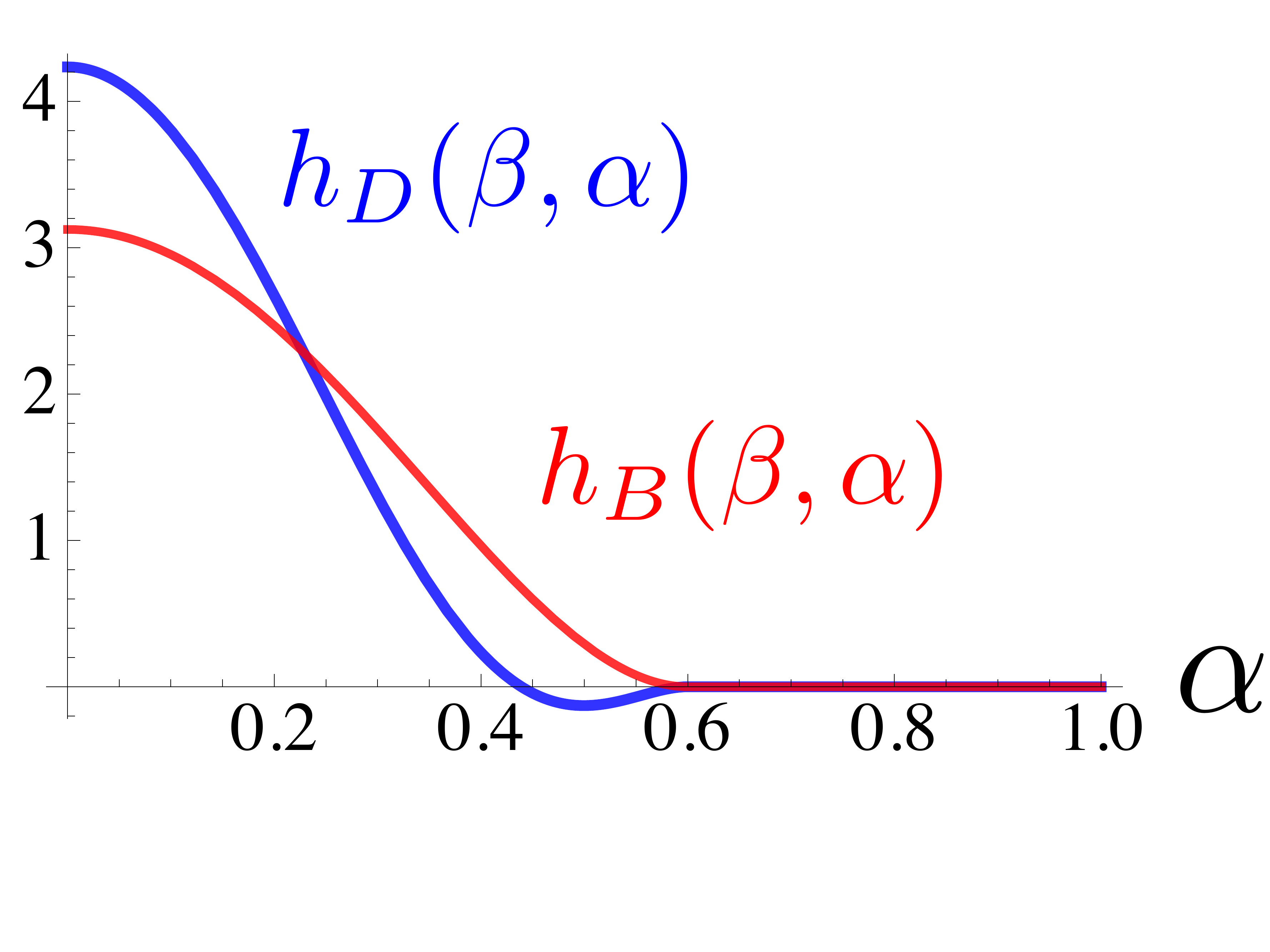} 
 \caption{Normalized profile functions  $h_B (\beta,\alpha) $ (red online) and 
 $h_D (\beta,\alpha) $ (blue online) described in the text, for \mbox{positive $\alpha$ }
 and 
 $\beta =0.2$ (top),  $\beta =0.4$ (bottom).}
 \label{hD}
 \end{center}
 \end{figure}

   By writing  ${\cal P}_{D} (\beta,\alpha) =  - e(\beta)h_D (\beta,\alpha) $
   one may introduce the profile function $h_D (\beta,\alpha) $ ``generating''
   DD  ${\cal P}_{D} (\beta,\alpha)$ from  the forward distribution $e(\beta)$:
       \begin{align}
   h_D (\beta,\alpha)= h_B (\beta,\alpha) + h_{\rm add} (\beta,\alpha) \  . 
   \end{align}   
   In addition to   the profile   term $h_B (\beta,\alpha) $  present  in ${\cal E}_{B} (\beta,\alpha)$,
it  also contains the term  $h_{\rm add} (\beta,\alpha)$ produced by the gauge transformation. It  
     satisfies  
    \begin{align}
  \int_{-1+|\beta|}^{1- |\beta|}  h_{\rm add} (\beta,\alpha) d \alpha  = 0 \  ,
   \end{align} 
    so that the total profile $h_D (\beta,\alpha) $ is still normalized to 1. 
   This  term  depends both on the shape of the initial profile $h_B (\beta,\alpha) $
   and on the form of the forward distribution $e(\beta)$.
   Being an even function of $\alpha$, the additional profile  function
   $h_{\rm add} (\beta,\alpha)$  cannot be positive definite.
   In fact, the total profile  $h_D (\beta,\alpha) $, in general,  is also not positive definite,
   see Fig.\ref{hD} for illustration referring to $N=2$ profile and forward function
   $e(\beta)= (1-\beta )^5/\sqrt{\beta}$.

    %     \begin{widetext}   
      
 %\end{widetext}   

   The representation (\ref{PDplusD}) corresponds to the  ``DD+D''  
   modeling. Usually, in this approach some positive definite 
   model profiles of Eq. (\ref{modn})  type   are 
   used to produce GPDs. Clearly, if one takes one of these 
   profiles, the result for $P_D(x,\xi)$ will be different from that
   obtained using DD ${\cal P}_{D} (\beta,\alpha)$ of Eq. (\ref{PD}). 
      This is an illustration of  the fact that  the 
      choices of  a profile {\" A}nsatze  in different gauges should be correlated.
      In particular, using the same profile $h(\beta,\alpha)$ 
      for ${\cal B}_{D} (\beta,\alpha)$  and  ${\cal P}_{D} (\beta,\alpha)/\beta $ 
      would produce different results for GPDs.

      \subsubsection{Model for $u$-quark border functions} 
      
     Let us consider a model with ``realistic'' assumptions.
     Take the valence part of $u$-quark distribution in the proton.
     For the usual parton distribution $f_u (x)$, we will take  the model 
         \begin{align}
 f_u (x) = \frac{5 \cdot 7}{2^4} \frac{(1-x)^3}{\sqrt{x}}
   \end{align} 
      normalized by the number  of $u$-quarks 
      in the proton
             \begin{align}
 \int_0^1 f_u (x) \, dx = 2 \ .
   \end{align} 
      The forward limit $e_u (x)$ of the $u$-quark GPD $E_u (x,\xi)$ 
      is normalized 
          \begin{align}
 \int_0^1 e_u (x) \, dx = \kappa_u 
   \end{align}      
      to the $u$-quark contribution $\kappa_u$ into the proton 
      anomalous magnetic moment $\kappa_p$. It is given by 
      $\kappa_u = 2\kappa_p + \kappa_n$, where $\kappa_n$ is the 
anomalous magnetic moment  of the neutron.
Numerically, $\kappa_u=1.673 $.  The model function
    \begin{align}
 e_u (x) = \frac{7 \cdot 9\cdot 11}{2^9} \kappa_u \frac{(1-x)^5}{\sqrt{x}}
   \end{align} 
satisfies the normalization constraint, has the same Regge behavior 
$1/\sqrt{x}$ for small $x$ as $f_u(x)$,  and has $(1-x)^5$ behavior 
for $x \to 1$, as suggested by form factor fits \cite{Diehl:2004cx,Guidal:2004nd}.

\begin{figure}[h]
\begin{center} 
\includegraphics[scale=0.21]{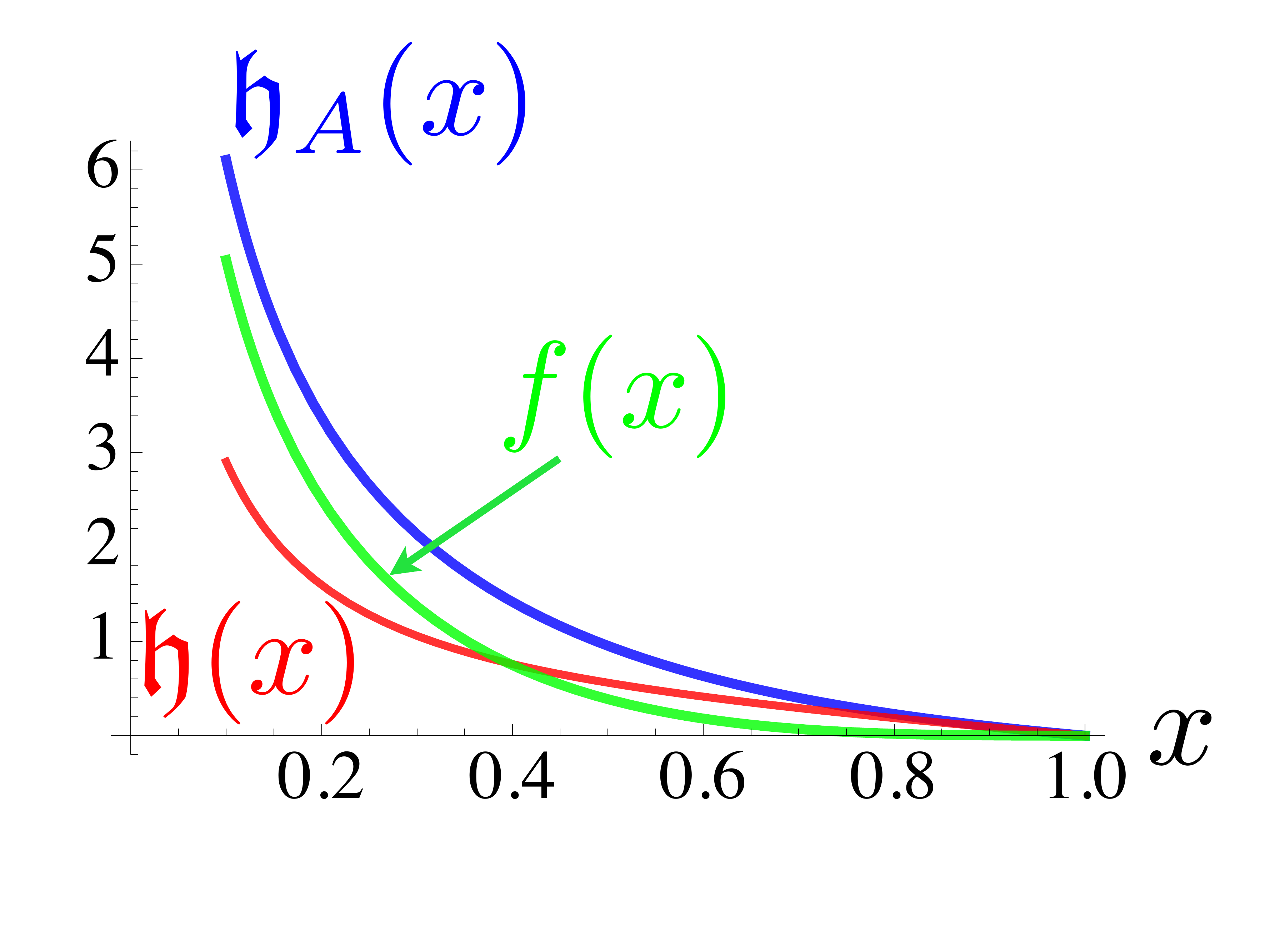}
%\vspace{-1cm}
\caption{Model functions for $u$-quark valence contributions described in the text:
usual parton density $f (x)$ (green online), border function  $ {\mathfrak h}_A (x)$
(blue online) and $ {\mathfrak h} (x)$
(red online).}
\label{hbord}
\end{center}
\end{figure}

For simplicity we take $N=1$ profile while building  $H$-GPD and $N=2$ in case of  $E$-GPD. 
Then, for the components of the border function 
    \begin{align}
 {\mathfrak h}_u (x) =  {\mathfrak h}_{A,u}^{(1)} (x) + {\mathfrak e}_{A,u} ^{(2)} (x) - {\mathfrak e}_{B,u}^{(2)}  (x)
   \end{align} 
we have 
  \begin{align}
  {\mathfrak h}_{A,u}^{(1)}  (x) = \frac{7}{2\sqrt{2}}  \frac{1- x }{(1+x)^{3/2}\sqrt{x}}  \  , 
   \end{align} 
  \begin{align}
  {\mathfrak e}_{A,u}^{(2)}  (x) = \frac{33 \kappa_u}{16\sqrt{2}}  \frac{(1- x)^2 }{(1+x)^{5/2}\sqrt{x}}  \  , 
   \end{align} 
and
\begin{align}
  {\mathfrak e}_{B,u}^{(2)}  (x) = \frac{99 \kappa_u}{32\sqrt{2}}  \frac{(1- x)^2 }{(1+x)^{3/2}\sqrt{x}}  \  
  .
   \end{align} 
The  ratio $  {\mathfrak e}_{B,u}^{(2)}  (x) /  {\mathfrak e}_{A,u}^{(2)}  (x)= \frac32 (1+x)$
is between 3/2 and 3 for $x$ changing from 0 to 1,
and the $ {\mathfrak e}$-addition  to the border function  $ {\mathfrak h}$ is negative. 
The resulting reduction of $ {\mathfrak h}$  is quite sizable, but 
the net result for $ {\mathfrak h}_u (x) $ is still positive, see Fig.\ref{hbord}.
One can see that in the middle region of $x$,   the border function  
$ {\mathfrak h}_u (x) $ in this particular model
 is rather close to the usual parton density $f_u (x)$.

Our goal here is  just to illustrate the basic principles 
of building models for the border functions,
so we will not analyze other possibilities,
like taking different profile functions for each ingredient 
of $ {\mathfrak h}_u (x) $, building models for 
$d$-quark distributions, etc.
Another feature that was not considered 
in this paper is the $t$-dependence of GPDs, 
which is very important in phenomenological applications. 
The  usual first step in this direction is to take a $t$-dependent
Regge parameter $a \to a(t)$, say, a linear Regge trajectory $a(t) = a(0) +\alpha' t$.
Similarly, for the D term contribution, one should take a $t$-dependent
form factor $\Delta (t)$
 (see Ref.\cite{Mezrag:2013mya} 
 for details).

 \section{Summary}

   Summarizing,
  in this  paper  we re-derived 
GPD sum rules  for  the  nucleon GPDs   $H$ and $E$
based on their    representation in terms of DDs ${\cal A}$ 
and ${\cal B}$ discussed in Ref. \cite{Radyushkin:2013hca}. 
 These sum rules, originally proposed in 
 \mbox{Refs. \cite{Teryaev:2005uj,Anikin:2007yh,Diehl:2007jb}} 
  allow to use border functions $ {\mathfrak h}(x)=H(x,x)$ and 
  \mbox{$ {\mathfrak e}(x)=E(x,x)$} 
  instead of full GPDs $H(x,\xi)$ and $E(x,\xi)$
  in the integrals producing (real part of) Compton form factors.
  The resulting description of DVCS
  (in $t=0$ limit)   in terms of functions $ {\mathfrak h}(x) $, 
  $ {\mathfrak e}(x)$
  depending on just one momentum fraction parameter $x$ is closer 
  to that of DIS that involves usual parton densities 
  $f(x)$. 
  
 Taking   formally  $\xi=0$  limit in 
  GPD sum rules produces integrals involving 
  the difference $ {\mathfrak h}(x) -f(x)$
  of border function  $ {\mathfrak h}(x) $ and parton density 
  $f(x)$.   However, both terms come in extremely singular 
      combinations ${\mathfrak h}(x) /x$ and $f(x)/x$ that 
      require subtraction prescription for  $x=0$ singularities. 
     Unfortunately, after implementation of the  subtraction 
     procedure, the resulting ``sum rule'' 
     contains no information about
      the difference between ${\mathfrak h}(x)$ and  $f(x)$
      for $x\neq 0$.

   To study the interrelation between  ${\mathfrak h}(x)$ and $f(x)$,  we used 
 models for GPDs based on 
 factorized DD Ansatz. In particular, we considered 
 a  model for the valence $u$-quark border function
 ${\mathfrak h}_u (x)$, and observed that 
 the function ${\mathfrak h}_{A,u}  (x)$ 
 constructed from $f_u(x)$  in a standard way 
 has a strong enhancement over $f_u(x)$. 
 However,  a more complicated ${\cal A} + {\cal B}$ 
  DD representation  for the nucleon GPDs \cite{Radyushkin:2013hca} 
requires an  extra term ${\mathfrak e}_{A,u}  (x) - {\mathfrak e}_{B,u}$
that considerably reduces the resulting  
border function ${\mathfrak h}_u (x)$  making it, in the middle region of $x$,
 rather close
to the parton density $f_u(x)$.
This observation illustrates the importance of taking into account 
the detailed structure of DD representations for nucleon GPDs. 

  While our discussion in the present paper 
  refers  to the formal $t=0$ limit,
  our results may be easily extended onto 
  $t$-dependent Compton form factors by  taking  a $t$-dependent
  Regge parameter $a \to a(t)$
  in models for input parton densities.

 \section*{Acknowledgements}
 
 I  thank 
     H. Moutarde 
  and C. Mezrag   for  discussions and  correspondence.

 This work is supported by Jefferson Science Associates,
  LLC under  U.S. DOE Contract No. DE-AC05-06OR23177.

  \bibliographystyle{apsrev4-1.bst}
\bibliography{nucleon0409.bib}

 \end{document}